\documentclass[notoc]{JHEP} 

\usepackage{epsfig}

\newcommand\fverb{\setbox\pippobox=\hbox\bgroup\verb}
\newcommand\fverbdo{\egroup\medskip\noindent%
			\fbox{\unhbox\pippobox}\ }
\newcommand\fverbit{\egroup\item[\fbox{\unhbox\pippobox}]}
\newcommand{\ppm}{$\pm$}
\newbox\pippobox

\title{Multiplicity distributions in $e^+e^-$ annihilation into hadrons
and pure birth branching processes}

\author{by O.G. Tchikilev\\
	Institute for High Energy Physics, 142284 Protvino, Russia\\
	E-mail: \email{tchikilov@mx.ihep.su}}

\preprint{\hepph{9912xxx}}	

\abstract{ 
 Recursive solution for a general homogeneous in time pure birth branching
 process with simultaneous production of any number of ``particles'' and with
 continuous evolution parameter is given. Calculational 
 algorithm based on the use of Koenigs function and functional
 Schr\"oder equation is
 described. It is shown that multiplicity distributions in $e^+e^-$ 
 annihilation
 into hadrons for c.m. energies up to 189~GeV are well described by the
 modified negative binomial distribution, explained by simple pure birth
 branching process without multiple simultaneous ``particle'' production.
 The energy dependence of the evolution parameter is also discussed.}

\begin{document} 

\maketitle 

\section{Introduction}

Recently it has been shown~[1-13] 
that multiplicity distributions in
 $e^+e^-$ annihilation into hadrons are fairly well described by the
modified negative binomial  distribution (MNBD). Two scenarios based on the
use of the pure birth branching processes 
(for description of the mathematical
formalism see for example~\cite{seva,athreya}) have been proposed to explain the
origin of the MNBD. In the scenario advocated in the paper~\cite{suzu1} the
binomial distribution of particle production sources occurs at some initial
stage, each source evolves afterwards according to the pure birth branching
process with immigration. In other scenario~\cite{chliap2,chliap3} 
a fixed number of particle production
sources occurs initially and they evolve also according to the pure birth 
branching process with production of  intermediate neutral clusters. The 
final state hadrons are produced through decay of these clusters. In the
 branching process used in both scenarios no multiple simultaneous 
 ``particle'' or cluster production is allowed.  In this letter we study the
 possibility of having  simultaneous multiple ``particle'' production. 
In  section~2  a recursive solution for the pure birth branching 
 process with 
 simultaneous production of any number of ``particles'' is 
 given
 together with a
 calculational algorithm, based on the use of  Koenigs function 
 and functional Schr\"oder equation\cite{fatou} 
 (recent description can be found 
 in~\cite{valiron,blanchard}). In  section~3  results 
 of fits of the charged particle multiplicity distributions in
 $e^+e^-$ annihilation into hadrons at c.m. energies above
 20~GeV~[19-28] to the MNBD and
 to the distribution resulting from the branching process with
 additional simultaneous pair
 ``particle'' production are presented. Our conclusions and  discussion are
 given in last section.

\section{Recursive solution for a general pure birth branching process}

 Let us recall that the branching process with continuous evolution
 parameter $t$ is determined by  the differential probability densities
 $\alpha_i$ for the transition of one object, ``particle'', 
 into $i$ objects, ``particles''.
 All ``particles'' are assumed to evolve independently.
 For a homogeneous branching process
 the $\alpha_i$ do not depend on $t$ and   probabilities for transition
 $1 \rightarrow i$ 
 in an infinitely small interval $\Delta t$ are

\begin{equation}
 p_{1 \rightarrow i}(t, t+\Delta t) = \alpha _i \Delta t ~~~,~i=0,2,3,...
\label{eq:1}
\end{equation}

and for transition 1 to 1

\begin{equation}
  p_{1 \rightarrow 1} (t, t+\Delta t) = 1 - \alpha \Delta t
\label{eq:2}
\end{equation}

with

\begin{equation}
 \alpha = \sum_{i \neq 1}^{\infty} \alpha_i \qquad .
\label{eq:3}
\end{equation}
 For the pure birth branching process $\alpha_0 = 0$.

 The  probability distribution $p_i(t) \equiv p_{1 \rightarrow
 i} (0,t)$ for the process having  one particle at $t=0$ can be found
from the forward  Kolmogorov equation

\begin{equation}
 \frac{\partial m}{\partial t} = 
 f(x)\frac{\partial m}{\partial x}\, \mbox{~~~with initial condition}
 \qquad m|_{t=0} = x \qquad ,
\label{eq:303}
\end{equation}
 for the probability generating function

\begin{equation}
 m(x,t) = \sum_{i=1}^{\infty} p_i (t) x^i \qquad ,
\label{eq:4}
\end{equation}
 with

\begin{equation}
 f(x) = \sum_{i=2}^{\infty} \alpha_i x^i - \alpha x \qquad .
\label{eq:5}
\end{equation}
 Using the Taylor expansion  of the
 equation~\ref{eq:303} over $x$ one can obtain the following system of
 differential equations  for the probabilities $p_{i}$

\begin{eqnarray}
 \frac{dp_1}{dt} = - \alpha p_1  \, \qquad ,\\
 \frac{dp_2}{dt} = \alpha_2 p_1 - 2 \alpha p_2\, \qquad , \\
 \frac{dp_3}{dt} = \alpha_3 p_1 + 2 \alpha_2 p_2 - 3 \alpha p_3\, \\
\label{eq:8}
\end{eqnarray}
   and for arbitrary $k$

\begin{equation}
 \frac{dp_k}{dt} =  \sum_{j=1}^{k-1} j \alpha_{k-j+1} p_j  
 - k \alpha p_k
\label{eq:87}
\end{equation}
with initial condition
\begin{equation}
  p_k(0) = \delta_{1k} \qquad .
\label{eq:88}
\end{equation}

These equations  can be solved one after another, for $p_1$

\begin{equation}
 p_1 = \exp {(-\alpha t)} \qquad ,
\label{eq:9}
\end{equation}
 for $p_2$

\begin{equation}
 p_2 = a_2 p_1 (1-p_1) \qquad ,
\label{eq:10}
\end{equation}
 where $a_i = \alpha_i/\alpha$ , and so on.

 One can  observe that $p_k$ has polynomial dependence of $k$-th power on
 $p_1$ 

\begin{equation}
 p_k = \sum_{i=1}^{k} \pi_{ik} p_{1}^{i}
\label{eq:12}
\end{equation}
 with constant coefficients $\pi_{ik}$. These coefficients can be found using
 the following recursions 

\begin{equation}
 \pi_{11} =1 \, , \pi_{12} = -\pi_{22} = a_2\, ,
\label{eq:13}
\end{equation}

\begin{equation}
  (k-i) \pi_{ik} = \sum_{j=1}^{k-1} j a_{(k+1-j)} \pi_{ij} =
  \sum_{j=1}^{k-1} ( k - j ) a_{(j+1)} \pi_{i(k-j)} 
~~~, i=1, 2, ..., k-1~ , 
\label{eq:14}
\end{equation}
 and the coefficient $\pi_{kk}$ can be found from the initial 
 condition~\ref{eq:88}

\begin{equation}
 \pi_{kk} = - \sum_{i=1}^{k-1} \pi_{ik} \qquad .
\label{eq:15}
\end{equation}

 This  solution satisfies also the backward Kolmogorov equation
 $dm/dt=f(m)$. Solving the backward equation in quadratures 
one can obtain a
 functional relation between  $m$ and $x$. For example, for the branching
 process with non-zero $\alpha_2$ and $\alpha_3$ one has:
\begin{equation}
 \varphi (m) = \varphi (x) \exp (-\alpha  t )
\label{eq:151}
\end{equation}
with
\begin{equation}
 \varphi (x) = \frac{x}{(1-x)^{\frac{\gamma}{\gamma+1}}
  ( x + \gamma)^{\frac{1}{\gamma+1}}}
\label{eq:152}
\end{equation}
and
\begin{equation}
  \gamma = \frac{\alpha_2 + \alpha_3}{\alpha_3} =
            \frac{\alpha}{\alpha_3}            \qquad.  
\label{eq:153}
\end{equation}

 In practical calculations the polynomial formulae \ref{eq:12} for $p_k$ 
 have limited  numerical precision. More robust calculational 
 algorithm based on the
 mathematical formalism used by Fatou\cite{fatou} in the studies of the 
 iterations of  rational functions is desribed briefly below. Let us denote

\begin{equation}
 m_n(x,t) = m(x,nt) \qquad .
\label{eq:16}
\end{equation}
 The following  relations are valid for the probability generating
functions $m_n(x,t)$ (see for example~\cite{seva,athreya})

\begin{equation}
 m_n(x,t) = m_{n-1} (m(x,t),t) = m(m_{n-1}(x,t),t) \qquad .
\label{eq:17}
\end{equation}
 In analogy with \cite{fatou} one can introduce the
  Koenigs function for the iterations~\ref{eq:17} as a limit:

\begin{equation}
  K(x) = \lim_{n \to \infty} \frac{m_n(x,t)}{p_1^n} \qquad .
\label{eq:18}
\end{equation}
 For the pure birth branching process  the Koenigs function has the following
 form
\begin{equation}
 K(x) = \sum_{i=1}^{\infty} \pi_{1i} x^i \qquad .
\label{eq:19}
\end{equation}
 Let us denote $\pi_{1i} = \kappa_{i}$.
 The recurrence relations  \ref{eq:14} lead to the 
following recurrence  for $\kappa_i$

\begin{equation}
 (i-1)\kappa_i = \sum_{j=1}^{i-1}  (i-j) a_{j+1} \kappa_{i-j}
 = \sum_{j=1}^{i-1} j a_{(i-j+1)} \kappa_{j} \qquad .
\label{eq:20}
\end{equation}
 The remarkable property of the Koenigs function\cite{fatou,valiron,blanchard}
 is  the functional Schr\"oder equation:

\begin{equation}
  K(m(x,t)) = p_1 K(x) \, ~~~~,
\label{eq:23}
\end{equation}
 it follows as a limit from the functional 
relations~\ref{eq:17}\footnote{ It is worth to note that the Schr\"oder equation
and the equation \ref{eq:151} are identical, therefore the Koenigs function
 for the branching
process with non-zero $\alpha_2$ and $\alpha_3$ coincides with
$\varphi(x)$ defined in \ref{eq:152}. In general case the $K(x)$ is 
proportional to $\exp (-\alpha \int \limits^{x} \frac{du}{f(u)})$.}.
 Let us denote

\begin{equation}
  m^j = \sum_{i=j}^{\infty} p_{i}^{(j)} x^j \qquad , j=2, 3, \cdots
\label{eq:24}
\end{equation}
 The use of the Taylor expansion of the Schr\"oder equation \ref{eq:23} 
 results in the following relations:

\begin{equation}
 p_1 \kappa_i = \sum_{j=1}^{i} \kappa_j p_{i}^{(j)} ~~~~,\qquad
 i=1, 2, 3, ...
\label{eq:25}
\end{equation}
 And $p^{(j)}_{i}$ can be found using successively the convolutions:

\begin{equation}
 p_j^{(2)} = \sum_{i=1}^{j-1} p_i p_{i-j} \qquad j=2,3, \ldots \\
\label{eq:261}
\end{equation}
\begin{equation}
 p_j^{(N)} = \sum_{i=1}^{j-1} p_i p_{j-1}^{N-1} \qquad j=N, N+1, \ldots
 \qquad N=3, 4, \ldots
\label{eq:262}
\end{equation}
  Using \ref{eq:25}, \ref{eq:261} and \ref{eq:262} one can obtain 
the recurrence for
$p_{n}$. Indeed, let us assume that we know $p_i$ for $i=1,2, \ldots, n-1$.
Then at step $n$ we calculate

\begin{equation}
 p_n^{(j)} = \sum_{i=1}^{n-j-1}p_i p_{n-i}^{(j-1)} \qquad j=2, \ldots, n-1
\label{eq:27}
\end{equation}
and

\begin{equation}
 p_n^{(n)} = p_1^n \qquad .
\label{eq:28}
\end{equation}
 Finally, the equation \ref{eq:25} leads to the following expression 

\begin{equation}
 p_n = \kappa_n (p_1 -p_1^n) - \sum_{j=2}^{n-1} \kappa_j p_n^{(j)} \qquad .
\label{eq:29}
\end{equation}

 The probability to produce $n_-$  negatively charged particles
($n_{-} = n_{ch}/2$)
in the scenario~\cite{chliap2,chliap3} can be found using the formula:

\begin{equation}
 P_{n_-} = \sum_{j=N}^{\infty} p_{j}^{(N)} q_{n_-}^{(j)}  =
 \sum_{j=N}^{\infty} p_{j}^{(N)} \frac{j!}{n_-!(j-n_-)!}
 \varepsilon ^{n_-} (1 - \varepsilon)^{j - n_-}  \qquad .
\label{eq:30}
\end{equation}
 Here $N$ denotes the number of initial sources of particle production, 
 $q_{n_-}^{(j)}$ is the usual binomial probability to have 
 $n_{-}$ 
 negatively charged particles
  ( we remind that $n_{-}$ in this scenario is equal to the number of
 produced charged particle pairs) from
 decay of $j$ clusters and
 $\varepsilon$ is the probability of cluster decay into pair of charged
 particles. In practical calculations we stop summation in \ref{eq:30} when
 the difference between $1$ and the sum of $p^{(N)}_{j}$ becomes smaller than
 the computer precision.
\FIGURE{\epsfig{file=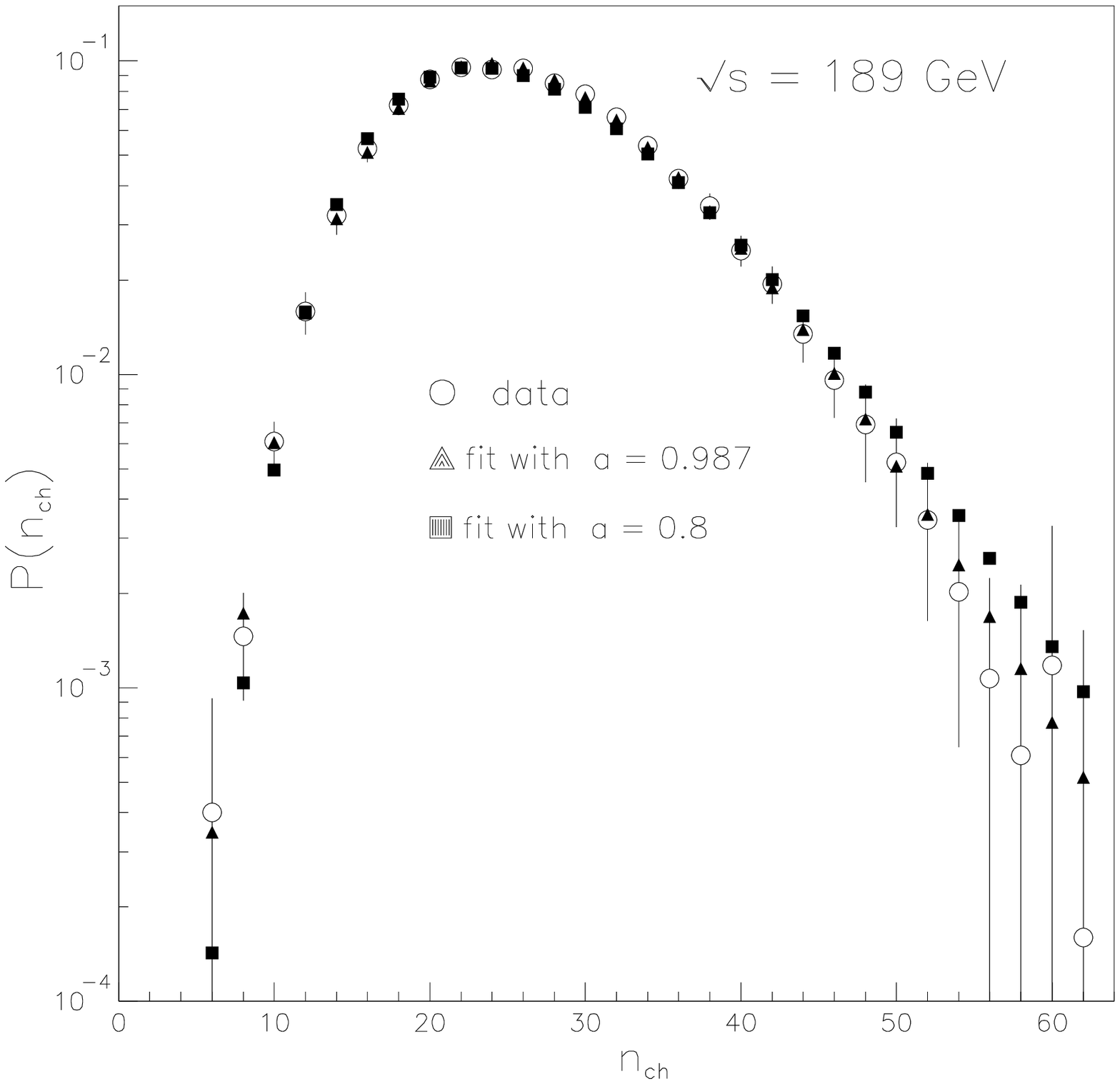,width=16cm}
\caption{Multiplicity distribution measured by the OPAL Collaboration at
 $\sqrt {s} = 189$~GeV compared with the predictions  of the fits with 
 $a = 0.987$ and $a=0.8$.}%
\label{opall}} 

\section{Results of fits}

 The pure birth branching process with non-zero $\alpha_2$ and $\alpha_3$ is
  determined by these differential probability densities
 and by the evolution parameter $t$.  The
  probability distribution $p_i(t)$ depends on some combinations of the
 $\alpha_2$ and $\alpha_3$ and on $p_1(t)$ (\ref{eq:9}).  
 The final multiplicity distribution (\ref{eq:30}) is characterized also by the
 number of initial sources $N$ and by the probability $\varepsilon$. 
 As in the papers~\cite{chliap3,ot1} we fix $N$ at the value $7$. We use as 
 free parameters in the fits  $p_1$, $\varepsilon$ and the ratio
 $a= \alpha_2/(\alpha_2 + \alpha_3)$. The parameter $a$ is equal to one
 for the MNBD and is equal to zero for the branching process with simultaneous
 pair ``particle'' production. Results of the fits are given in 
 Table~1. In the cases when the three-parameter fit has minimum with $a=1$, 
 only the  results of the two-parameter fit with fixed $a=1$ are shown.

 The influence of the parameter $a$ is illustrated in  Figure~\ref{opall},
 where the multiplicity distribution at $\sqrt{s} = 189$~GeV is compared with
 the predictions of the three-parameter fit  and with
 the predictions of the two-parameter fit with fixed $a=0.8$, having 
  $p_1 = 0.509 \pm 0.007$, $\varepsilon = 0.861 \pm 0.012$ and
  $\chi^2/NDF = 19.1/(29-2)$. From Figure~\ref{opall} it is seen that 
the predictions of the fit with $a=0.8$ exceed the predictions of the 
three-parameter fit at $n_{ch}$ below $12$, the opposite is observed
   at $n_{ch}$ above $40$.
   
\TABLE{\begin{tabular}{|c|c|c|c|c|c|c|}
\hline
   exp. & $\sqrt{s}$ & $p_{1}$ & $a$ & $\varepsilon$ & $\chi^{2}$ & $N_{p}$ \\
\hline
       TASSO~[19] & 22  & 0.773\ppm  0.008 & 1
 & 0.626\ppm 0.007 & 22.1 & 14 \\ 
        &   & 0.854\ppm  0.025 & 0.706\ppm  0.138 
 & 0.658\ppm  0.138 & 11.4 &  \\
  HRS~[20]        & 29  & 0.758\ppm  0.006 & 1           & 0.697\ppm  0.006 
 & 7.8  & 14 \\
 TASSO~[19]       & 34.8& 0.688\ppm  0.004 & 1           & 0.668\ppm  0.004
 & 15.5 & 18 \\
        & & 0.714\ppm  0.011 & 0.948\ppm  0.024 
 & 0.681\ppm  0.007 & 9.8  &  \\
 TASSO~[19]       & 43.6& 0.645\ppm  0.006 & 1           & 0.693\ppm  0.007 
 & 27.1 & 19 \\
  AMY~[21]        & 50  & 0.658\ppm  0.018 & 1   & 0.766\ppm  0.017
 & 2.3 & 19 \\
          &   & 0.674\ppm  0.068 & 0.973\ppm  0.112 
 & 0.772\ppm  0.029 & 2.2  &  \\
             & 52  & 0.634\ppm  0.014 & 1           & 0.753\ppm  0.015 
 & 5.8  & 19 \\
             & 55  & 0.637\ppm  0.015 & 1  & 0.767\ppm 0.015 & 4.6& 19\\
             &   & 0.650\ppm  0.042 & 0.975\ppm  0.081 & 0.772\ppm  0.022 
 & 4.5  &  \\
             & 56  & 0.623\ppm  0.012 & 1           & 0.757\ppm  0.014 
 & 11.0 & 19 \\
             & 57& 0.614\ppm 0.014 & 1 & 0.767\ppm 0.014& 7.5& 19\\
             &   & 0.616\ppm  0.038 & 0.994\ppm  0.070 & 0.768\ppm  0.021 
 & 7.5  &  \\
             & 60  & 0.595\ppm  0.016 & 1           & 0.749\ppm  0.017 
 & 5.8  & 20 \\
             & 60.8& 0.644\ppm  0.012 & 1           & 0.791\ppm  0.012 
 & 14.4 & 20 \\
             & 61.4& 0.657\ppm  0.036 & 0.935\ppm  0.075 & 0.793\ppm  0.018 
 & 9.4  & 20 \\
 ALEPH~[22]      & 91  & 0.507\ppm  0.011 & 1           & 0.757\ppm  0.017 
 & 7.4  & 26 \\
 DELPHI~[23]     & 91  & 0.512\ppm  0.003 & 1           & 0.774\ppm  0.004 
 & 30.1 & 25 \\
   L3~[24]        & 91  & 0.512\ppm  0.010 & 1           & 0.751\ppm  0.014 
 & 11.8 & 23 \\
 OPAL~[25]        & 91  & 0.501\ppm  0.005 & 1           & 0.776\ppm  0.008 
 & 5.0  & 25 \\
 OPAL~[26]        & 133& 0.456\ppm 0.019 & 1 & 0.762\ppm 0.031 & 5.3 & 25\\
         &  & 0.525\ppm  0.059 & 0.889\ppm 0.106 & 0.821\ppm  0.054 
 & 3.6  &  \\
 OPAL~[27]        & 161 & 0.441\ppm  0.017 & 1           & 0.768\ppm  0.031 
 & 3.6  & 25 \\
 OPAL~[28]   & 172 & 0.463\ppm 0.019 & 1 &  0.838\ppm 0.034 & 5.8 & 29 \\
 OPAL~[28]   & 183 & 0.437\ppm 0.012 & 1 & 0.826\ppm 0.020 & 13.8 & 29 \\
    & & 0.496\ppm 0.034 & 0.904\ppm 0.060 & 0.882\ppm 0.033 & 10.3 & \\
 OPAL~[28]  & 189 & 0.402\ppm 0.008& 1 & 0.772\ppm 0.014 & 3.3 & 29 \\
    & & 0.411\ppm 0.025 & 0.987\ppm 0.037 & 0.780\ppm 0.026 & 3.1 & \\    
\hline
\end{tabular}
\caption{Results of the fits of the distribution \ref{eq:30} to the
 charged particle multiplicity distributions in $e^+e^-$ annihilation
into hadrons.}\label{tab:mul}}

  One can see from  Table~1 that the quality of the fits is quite good 
 and that the values of $\varepsilon$ and $\chi^2$ for the
MNBD fits are practically the same as in the previous 
publications~\cite{chliap1,chliap2,chliap3}.
 The values of the parameter $a$ are concentrated near one and the allowed
fraction for simultaneous pair production is below a few per cent. We have
checked also that the parameter $a$ is  near one in the fits with $N$ different
from seven, in most cases these fits have higher $\chi^2$ than 
the fits with $N=7$ (see also \cite{chliap3}).
\FIGURE{\epsfig{file=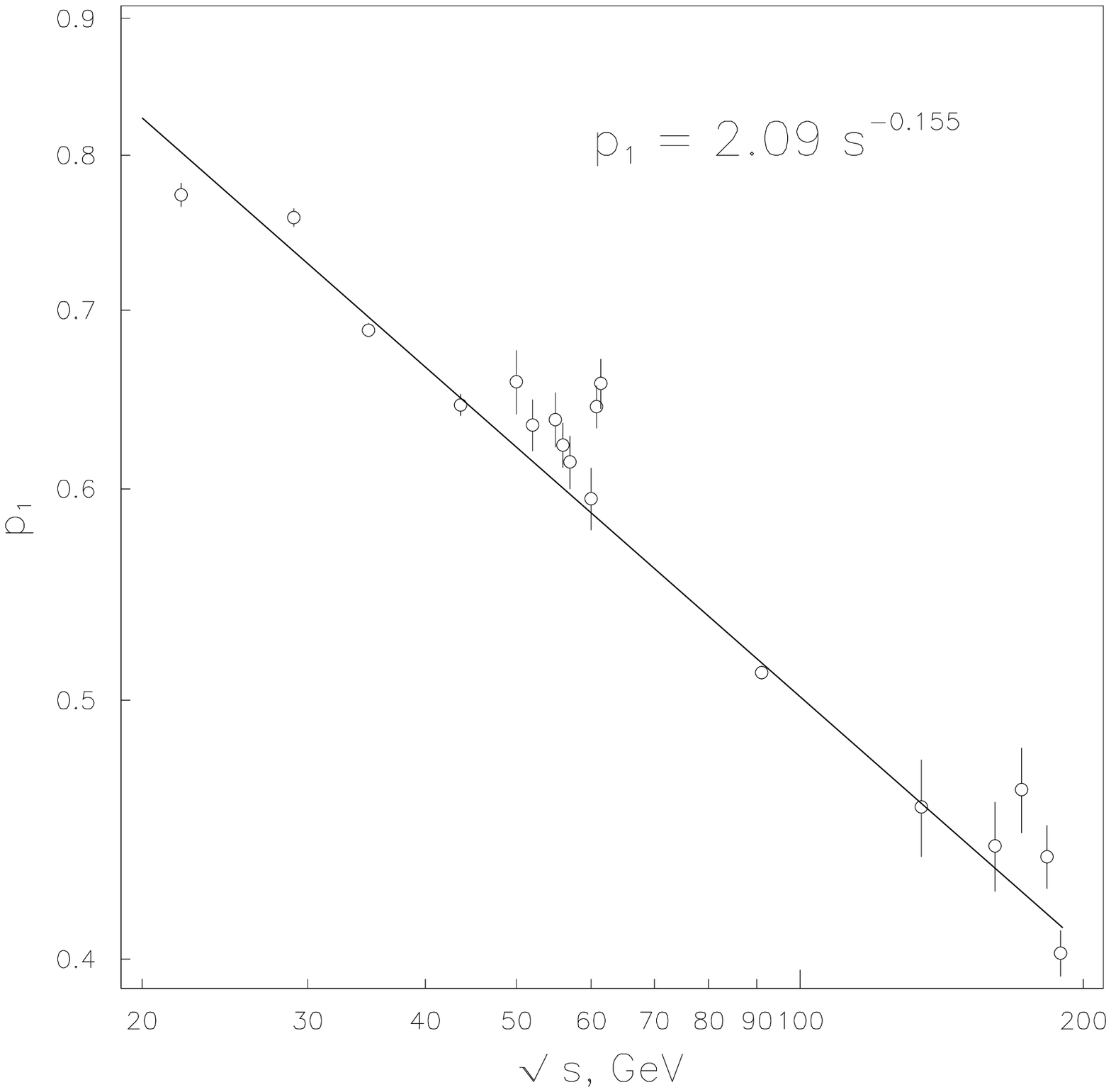,width=16cm} 
        \caption{Energy dependence of the probability $p_1$ for
 $e^+e^-$ annihilation into hadrons. Solid line shows the predictions of the
 power law fit.}%
	\label{myfig}}

\section{Discussion and conclusions}

 In recent publication~\cite{biya} Biyajima and coworkers have observed
that the LEP data at $\sqrt {s} = 133$~GeV~\cite{opal2} favour 
logarithmic dependence $\sim \log{\sqrt{s}}$ of the evolution parameter $t$ in
contradiction with $\log{\log{Q^2}}$ prediction given by QCD. For logarithmic
dependence the probability $p_1$ should have power law dependence on the c.m. 
energy $\sqrt{s}$. The energy dependence of the parameter $p_1$ for the
MNBD fits is shown in  Figure~\ref{myfig}.  From  
Figure~\ref{myfig} it is seen that 
recent LEP measurements at 161~\cite{opal3}  172, 183 and 
189~\cite{opal4}~GeV also favour the logarithmic dependence of the evolution
parameter on $\sqrt{s}$. The data 
are fairly well described by the power law dependence 
 $p_1 = A (\sqrt{s})^{-B}$ with the slope $B = 0.310 \pm 0.005$ and with
 $\chi^2 /NDF = 104.1/(21-2)$.  As noted in~\cite{biya}, this 
contradiction with QCD is probably attributed to hadronization effects. 

 Our conclusions are the following. The recursive solution for the general
pure birth branching process is given. The calculational algorithm based on
the use of the Koenigs function and the functional Schr\"oder equation is
described. The results of the fits to the  charged particle multiplicity
distributions in $e^+e^-$ annihilation into hadrons for c.m. energies up to
189~GeV show the validity of the MNBD parametrization and the absence of the
component with simultaneous pair production at the level exceeding a few 
per cent.
The latest LEP measurements favour the logarithmic dependence of the
evolution parameter $t$ as noted earlier\cite{biya}.


\acknowledgments

 I am greatly indebted to Yu.~G.~Stroganov for the hints how to find the
recursive solution for the general pure birth branching process.
I am grateful to M.~Yu.~Bogolyubsky
for reading of the manuscript and useful remarks.
\bigskip


\end{document}